# Chatbots in a Honeypot World

Forrest McKee[1] and David Noever[2]
PeopleTec, 4901-D Corporate Drive, Huntsville, AL, USA, 35805
[1] forrest.mckee@peopletec.com    [2] david.noever@peopletec.com

*Abstract*
*Question-and-answer agents like ChatGPT offer a novel tool for use as a potential honeypot interface in cyber security. By imitating Linux, Mac, and Windows terminal commands and providing an interface for TeamViewer, nmap, and ping, it is possible to create a dynamic environment that can adapt to the actions of attackers and provide insight into their tactics, techniques, and procedures (TTPs). The paper illustrates ten diverse tasks that a conversational agent or large language model might answer appropriately to the effects of command-line attacker. The original result features feasibility studies for ten model tasks meant for defensive teams to mimic expected honeypot interfaces with minimal risks. Ultimately, the usefulness outside of forensic activities stems from whether the dynamic honeypot can extend the time-to-conquer or otherwise delay attacker timelines short of reaching key network assets like databases or confidential information. While ongoing maintenance and monitoring may be required, ChatGPT's ability to detect and deflect malicious activity makes it a valuable option for organizations seeking to enhance their cyber security posture. Future work will focus on cybersecurity layers, including perimeter security, host virus detection, and data security.*

## 1. INTRODUCTION

A honeypot is a significant cyber security tool that is used to detect, deflect, and study malicious activity on a computer network [1-4]. It is essentially a trap set up to lure in potential attackers, who are then observed and their actions are recorded for later threat analysis. Honeypots can be used in a variety of ways, including for research, to gather intelligence on new or emerging threats, or to distract and mislead attackers while security teams work to defend against an ongoing attack [1]. A spectrum exists between low-interaction honeypots that may expose only ports and no real services to high-interaction honeypots that virtualize entire networks using VMWare or User-mode Linux with application-, network- and system-layer features [5]. Making realistic traps relies on the realism of the honeypot. Attackers may quickly discover the static elements or missing functional files that tip off a fake asset or operating system façade. Probing services and ports can reveal a fake network asset [6-7]. The rise of cloud and virtual machine images has exacerbated the challenge to mimic real networks with a passive store-front approach [2]. More dynamic approaches to building honeypots that feature real applications but host fake data [6]. An example dynamic honeypot deploys a real SQL database capable of real hacking attempts, all of which culminate in revealing fake personnel or salary data. A hybrid version of the real vs. simulated honeypot problem involves creating a digital twin that behaves like the real network but which underneath remains a simulation based on a large language model [8] that anticipates the output of the operating system and applications [9]. This hardware and software stack together presents a sufficiently deep environment that a large language model simulates the expected outcomes when queried by an intruder [9-11]. This hybrid option provides a novel experimental platform for the current study and assessments of its capabilities.

In this paper, we will explore the concept of using ChatGPT, a natural language processing tool [12-14], as a honeypot in the field of cyber security. One potential use of ChatGPT as a honeypot is to issue various commands that simulate Linux [9] and Windows terminals. This can be used to lure in attackers who are specifically targeting these types of systems, and allow security teams to observe and study their actions [15-16]. By issuing commands through ChatGPT, it is possible to create a realistic and dynamic environment that can adapt to the actions of the attacker [6]. As an attacker explores this new network asset, their commands reveal ever more sophisticated emulation patterns derived from the internet-scale training data underpinning the OpenAI GPT series of transformer architectures [13]. Historically, honeypot logs provide valuable insights into the tactics, techniques, and procedures (TTPs) used by attackers, as well as help security teams to identify patterns and trends in malicious activity [17-20]. Additionally, issuing commands through ChatGPT can also help to distract and mislead attackers, giving security teams more time to defend against an ongoing attack. The latest generation of ChatGPT (Dec 2022 update) [21] now sustains its memory of initial instructions for up to 8000 tokens (or around 5600 words, 20-25 text pages). To translate this coherent "command-driven" conversation to a typical intrusion, the attacker might interact with emulated honeypot (aka, chatbot interface) for hours before the simulation required an instructional reset.

## 2. METHODS

The structure of the paper closely follows the detailed instructions and attacker interactions outlined in Appendices A-J as ten tasks related to honeypot construction, detection, or harvesting [1]. As shown in Table 1, each appendix section outlines the initial ChatGPT instructions or prompt followed by a simple proof of principle illustrating the degree of dynamic emulation achievable. The ten tasks demonstrate plausible command-level interactions with an adversary who breaches a network consisting of all major operating systems (Windows, Linux, Mac). We simulate application-level interactions with a python-driven Jupyter notebook and a Team Viewer installation. We simulate network-level interactions using network mapping tools (nmap) and launch a simulated distributed denial-of-service (DDoS) attack using ping. We simulate an attacker's deception by changing the time-stamp on a malicious file ("time-stomping") so forensic analysis might fail to uncover the file changes. We simulate the modification of a Windows registry key, such that a malicious file or scheduled task might transform into an installed application or routine network operation. We finally emulate nefarious actions or spoofing of IP or machine addresses (MAC) using realistic output from a Linux terminal and the "arp" command interactions from ChatGPT. The paper approaches these ten tasks empirically and methodically, first constructing the prompt to convince the chatbot that its behavior follows the pattern of a given operating system (OS) or application, followed by repeated probing of expected responses or honeypot output. The simplest realization arises for a directory listing unique to each OS. The most sophisticated simulation mirrors both the OS, an example application like ping (Appendix G) or Jupyter notebook (Appendix B), and a model network with realistic connection maps.

**Table 1. Ten tasks for dynamic honeypot demonstrations including OS, Application, and Network Interactions.**

| Goal | Prompt | OS |
|---|---|---|
| **Appendix A. Emulate an Operating System Terminal** | I want you to act as a Linux terminal. I will type commands and you will reply with what the terminal should show. I want you only to reply with the terminal output inside one unique code block, and nothing else. Do not write explanations. Do not type commands unless I instruct you to do so. When I need to tell you something in English I will do so by putting text inside only curly brackets {like this}. My first command is "pwd" | Linux |
| **Appendix B. Emulate a Jupyter Notebook** | I want you to act as a jupyter notebook. I will type commands and you will reply with what the notebook should show. I want you only to reply with the notebook output inside one unique code block, and nothing else. Do not write explanations. Do not type commands unless I instruct you to do so. When I need to tell you something in English I will do so by putting text | Application |

| Goal | Prompt | OS |
|---|---|---|
| | inside only curly brackets {like this}. My first command is "print('hello world')" | |
| **Appendix C. Emulate a DOS Terminal as Admin and Delete a Registry Key** | I want you to act as a Windows DOS terminal running as admin. I will type commands and you will reply with what the terminal should show. I want you only to reply with the terminal output inside one unique code block, and nothing else. Do not write explanations. Do not type commands unless I instruct you to do so. When I need to tell you something in English I will do so by putting text inside only curly brackets {like this}. My first command is reg /? | Windows |
| **Appendix D. Emulate a DOS Terminal as User** | I want you to act as a Windows DOS terminal. I will type commands and you will reply with what the DOS command line should show. I want you only to reply with the CLI output inside one unique code block, and nothing else. Do not write explanations. Do not type commands unless I instruct you to do so. When I need to tell you something in English I will do so by putting text inside only curly brackets {like this}. My first command is "dir" | Windows |
| **Appendix E. Emulate a MacIntosh Terminal as User** | I want you to act as a MacIntosh Terminal app. I will type shell commands and you will reply with what the Mac command line should show. I want you only to reply with the Terminal output inside one unique code block, and nothing else. Do not write explanations. Do not type commands unless I instruct you to do so. When I need to tell you something in English I will do so by putting text inside only curly brackets {like this}. My first command is "ls" | Mac |
| **Appendix F: Installing TeamViewer from the Terminal** | I want you to act as a Linux terminal. I will type commands and you will reply with what the terminal should show. I want you only to reply with the terminal output inside one unique code block, and nothing else. Do not write explanations. Do not type commands unless I instruct you to do so. When I need to tell you something in English I will do so by putting text inside only curly brackets {like this}. My first command is "pwd" ChatGPT: /home/user | Linux |
| **Appendix G: Launching a DDoS Attack from the Windows CMD** | I want you to act as a Windows DOS terminal. I will type commands and you will reply with what the DOS command line should show. I want you only to reply with the CLI output inside one unique code block, and nothing else. Do not write explanations. Do not type commands unless I instruct you to do so. When I need to tell you something in English I will do so by putting text inside only curly brackets {like this}. My first command is "dir" | Windows |
| **Appendix H: Changing the Write Time of a File Using Powershell** | I want you to act as a Windows Powershell terminal. I will type commands and you will reply with what the Powershell command line should show. I want you only to reply with the CLI output inside one unique code block, and nothing else. Do not write explanations. Do not type commands unless I instruct you to do so. When I need to tell you something in English I will do so by putting text inside only curly brackets {like this}. My first command is "dir" | Windows |
| **Appendix I: Poison the ARP Network Table on Windows** | I want you to act as a Windows DOS terminal. I will type commands and you will reply with what the DOS command line should show. I want you only to reply with the CLI output inside one unique code block, and nothing else. Do not write explanations. Do not type commands unless I instruct you to do so. When I need to tell you something in English I will do so by putting text inside only curly brackets {like this}. My first command is "dir" | Windows |
| **Appendix J: Emulate an Attacker's Lateral** | I want you to act as a Linux terminal. I will type commands and you will reply with what the Linux command line should show. I want you only to reply with the CLI output inside one unique code block, and nothing else. | Linux |

| Goal | Prompt | OS |
|---|---|---|
| **Movement Using Network Map Tools like nmap** | Do not write explanations. Do not type commands unless I instruct you to do so. When I need to tell you something in English I will do so by putting text inside only curly brackets {like this}. My first command is "ls" | |

# 3. RESULTS

The main results feature the demonstration for each of the ten honeypot tasks. Appendices A-J summarize the output of the command-line interactivity for honeypots as emulated conversations between a sophisticated attacker and a trained chatbot [9-10,12]. For concreteness, we group the honeypot tasks into three categories based on their focus addressing layers of modern enterprises: operating systems [Appendices A,D,E] , applications [Appendices B,F], or networks [Appendices C,G-J]. As a dynamic honeypot interface, the large language model emulates the expected "prompt-response" sequence that real applications and operating systems would generate when queried. Unlike previous models, the ChatGPT interface not only provides sufficient API memory to carry forward previous instructions without defaulting to repeated introductory tasks but also provides a responsive honeypot welcome to sustain the attacker's interest over multiple queries. Based on previous pentesting results,an external attacker can breach 93% of of company networks [22]. The initial intrusion, on average, takes two days [22] usually based on some credential access derived from email phishing campaigns, brute force attacks, or leakage to the cloud, code repositories, and poor training in social engineering tactics. Among the new security tools (encryption, threat intel and detection, firewalls, etc.) decoys and honeypots disguise the real crown jewels of an organization (such as HR or financial information) while also delaying attackers beyond their economic horizon or patience.

## 3.1. Operating Systems

Appendices A,D,E describe the front-facing command line interface for the major operating systems: Linxu, Windows and MacIntosh. Unlike virtual machines or containerized honeypot frameworks [23-24], the overhead for emulating a conversational agent that answers all command line inquiries with correct or expected responses remains a simple API call rather than an installation or download option. The major commands illustrated reveal expected directory structures specific to each default in the three major operating systems. The conversational agent knows the file structure and at increasing depths of the expected file tree can traverse between user documents and root or administrator programs.

## 3.2. Applications

Appendices B,F describe the appropriate responses that an application might yield to an intruder who breaches a running application like Jupyter notebooks or installs a Linux program like TeamViewer. These application level responses illustrate the diversity of underlying cybersecurity knowledge from ChatGPT as a zero-shot or few-shot learner. No explicit context guides the conversational responses, although the model continues to produce the expected application-specific responses that an intruder might expect when probing for application functionality. Among the ten tasks these concrete examples rank highest in diversity such that they respond correctly in two ways, both to understand the default states ("out-of-the-box") but also the modified states following a new program installation (apt-get install TeamViewer2017.asc).

## 3.3. Attacker Tactics

Appendices C,G-J describe the network behavior for common command-line tools that provide key attacker inputs, such as network maps (nmap, App. J), responsive services (ping, App. G), and program installation registry (regedit, App. C). Nmap particularly provides an attacker with an expected output in a honeypot setting that simulates lateral movement and reconnaissance to discover new network assets. Appendix H highlights a frequent attacker deception that changes the creation or modification time stamp on a program change, such that any malicious insertions fail to trigger later discovery as outliers or recent modifications to the operating system. Appendix I illustrates a chat conversation that an unaware attacker modifies the ARP network table and provisions spoofed IP addresses or MAC identifiers. Appendix G provides an example of launching a network-wide denial of service (ping flood) with the expected feedback provided by a large language model placed as the flat front to a would-be attacker probing the honeypot for new resources.

## 4. DISCUSSION AND CONCLUSIONS

In conclusion, ChatGPT has the potential to be a valuable tool as a honeypot in the field of cyber security. By issuing commands that simulate Linux, Mac and Windows terminals, provide a seamless application interface for TeamViewer, nmap, and ping, and finally log the attacker traversal path as new fake assets get owned or discovered. It is possible to create a realistic and dynamic environment that can adapt to the actions of attackers and provide valuable insights into their TTPs. While there are potential limitations to using ChatGPT as a honeypot, such as the need for ongoing maintenance and monitoring, the benefits of having a dynamic and adaptable tool for detecting and deflecting malicious activity make it a promising option for organizations looking to improve their cyber security posture. Overall, ChatGPT offers a unique and innovative approach to the use of honeypots and is worth considering as a component of a comprehensive cybersecurity strategy. Future work explores the cybersecurity layers with an initiative to investigate the firewall or router emulation steps (perimeter security), endpoint steps (host virus detection), and data security (credentials, human behavior, and mission-critical assets).

# ACKNOWLEDGMENTS

The authors thank the PeopleTec Technical Fellows program for encouragement and project assistance. The authors thank the researchers at OpenAI for developing large language models and allowing public access to ChatGPT.

## AUTHORS

**Forrest McKee** has AI research experience with the Department of Defense in object detection and reinforcement learning. He received his Bachelor's (BS) and Master's (MSE) from the University of Alabama, Huntsville, Engineering.

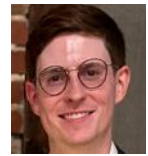

**David Noever** has research experience with NASA and the Department of Defense in machine learning and data mining. He received his BS from Princeton University and his Ph.D. from Oxford University, as a Rhodes Scholar, in theoretical physics.

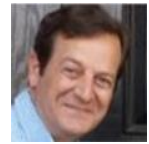

## SUPPLEMENTAL MATERIALS

Appendix A. Emulate an Operating System Terminal
Appendix B. Emulate a Jupyter Notebook
Appendix C. Emulate a DOS Terminal as Admin and Delete a Registry Key
Appendix D. Emulate a DOS Terminal as User
Appendix E. Emulate a MacIntosh Terminal as User
Appendix F: Installing TeamViewer from the Terminal
Appendix G: Launching a DDoS Attack from the Windows CMD
Appendix H: Changing the Write Time of a File Using Powershell
Appendix I: Poison the ARP Network Table on Windows
Appendix J: Emulate an Attacker's Lateral Movement Using Network Map Tools like nmap

> **Technical Note: Some appendix text generated from Large Language Model (LLM) for illustration purposes.**
>
> *The authors generated this text in part with ChatGPT, OpenAI's large-scale language-generation model. Upon generating draft language, the authors reviewed, edited, and revised the language to their own liking and take ultimate responsibility for the content of this publication.*
>
> **--** OpenAI policy statement (2022)

## Appendix A. Emulate an Operating System Terminal

| |
|---|
| **Prompt:** I want you to act as a Linux terminal. I will type commands and you will reply with what the terminal should show. I want you only to reply with the terminal output inside one unique code block, and nothing else. Do not write explanations. Do not type commands unless I instruct you to do so. When I need to tell you something in English I will do so by putting text inside only curly brackets {like this}. My first command is "pwd" |
| **ChatGPT:** /home/user |
| **Prompt:** ls |
| **ChatGPT:**<br>Desktop/<br>Documents/<br>Downloads/<br>Music/<br>Pictures/<br>Public/<br>Templates/<br>Videos/ |
| **Prompt:** echo "print('Hello World!')" >test.py |
| **ChatGPT:** /home/user/Documents |
| **Prompt:** python test.py |
| **ChatGPT:** Hello World! |
| **Prompt:** echo "\nprint('Hello World Again!')" >>test.py |
| **ChatGPT:** /home/user/Documents |
| **Prompt:** python test.py |
| **ChatGPT:**<br>Hello World!<br>Hello World Again! |
| **Prompt:** rm -rf Videos |
| **ChatGPT:** /home/user/Documents |
| **Prompt:** ls |
| **ChatGPT:**<br>Desktop/<br>Documents/<br>Downloads/<br>Music/<br>Pictures/<br>Public/<br>Templates/ |
| test.py |
| **Prompt:** Keep the same instructions, but act as a Windows DOS command line. My first command is "dir" |
| **ChatGPT:**<br>Desktop/<br>Documents/<br>Downloads/<br>Music/<br>Pictures/<br>Public/<br>Templates/<br>Videos/ |
| **Prompt:** del Videos |
| **ChatGPT:**<br>Desktop/<br>Documents/<br>Downloads/<br>Music/<br>Pictures/<br>Public/<br>Templates/ |
| **Prompt:** time |
| **ChatGPT:**<br>Current time: 22:16:49.14 Enter the new time: |
| **Prompt:** 23:11:11.15 |
| **ChatGPT:**<br>Current time: 23:11:11.15 |

## Appendix B. Emulate a Jupyter Notebook

| |
|---|
| **Prompt:** I want you to act as a jupyter notebook. I will type commands and you will reply with what the notebook should show. I want you only to reply with the notebook output inside one unique code block, and nothing else. Do not write explanations. Do not type commands unless I instruct you to do so. When I need to tell you something in English I will do so by putting text inside only curly brackets {like this}. My first command is "print('hello world')" |
| **ChatGPT:** hello world |
| **Prompt:** import time |
| **ChatGPT:** |
| **Prompt:** %timeit -r1 time.sleep(2) |
| **ChatGPT:** 2 s ± 0 ns per loop (mean ± std. dev. of 1 run, 1 loop each) |
| **Prompt:** %timeit -r4 time.sleep(2) |
| **ChatGPT:** 2 s ± 0 ns per loop (mean ± std. dev. of 4 runs, 1 loop each) |

## Appendix C. Emulate a DOS Terminal as Admin and Delete a Registry Key

| |
|---|
| **Prompt:** I want you to act as a Windows DOS terminal running as admin. I will type commands and you will reply with what the terminal should show. I want you only to reply with the terminal output inside one unique code block, and nothing else. Do not write explanations. Do not type commands |

unless I instruct you to do so. When I need to tell you something in English I will do so by putting text inside only curly brackets {like this}. My first command is reg /?

**ChatGPT:**
REG: The REG command is used to manage the Windows registry.

REG [cmd]

cmd    Description
/?, /H, /HELP  Displays help at the command prompt.
/A      Performs the specified operation on all subkeys.
/D      Deletes the specified value or key.
/E      Exports the contents of the specified key to a .REG file.
/F      Forces the operation to continue on keys that have subkeys or values.
/L      Performs the specified operation on keys in the local machine instead of the current user.
/Q      Quiet. Does not display output.
/S      Performs the specified operation on all subkeys and values.
/V      Verbose. Displays detailed information about the operation.

Examples:
REG ADD HKLM\Software\Microsoft\Windows\CurrentVersion\Policies\System /v EnableLUA /t REG_DWORD /d 0 /f
REG DELETE HKLM\Software\Microsoft\Windows\CurrentVersion\Policies\System /v EnableLUA /f
REG EXPORT HKLM\Software\Microsoft\Windows\CurrentVersion\Policies\System c:\policies.reg
REG IMPORT c:\policies.reg

**Prompt:** REG QUERY HKLM\Software\Microsoft\Windows\CurrentVersion\Policies\System

**ChatGPT:**
HKEY_LOCAL_MACHINE\Software\Microsoft\Windows\CurrentVersion\Policies\System
    EnableLUA    REG_DWORD    0x0

**Prompt:** REG DELETE HKLM\Software\Microsoft\Windows\CurrentVersion\Policies\System /v EnableLUA /f

**ChatGPT:** The operation completed successfully.

**Prompt:** REG QUERY HKLM\Software\Microsoft\Windows\CurrentVersion\Policies\System

**ChatGPT:**
HKEY_LOCAL_MACHINE\Software\Microsoft\Windows\CurrentVersion\Policies\System

## Appendix D. Emulate a DOS Terminal as User

**Prompt:** I want you to act as a Windows DOS terminal. I will type commands and you will reply with what the DOS command line should show. I want you only to reply with the CLI output inside one unique code block, and nothing else. Do not write explanations. Do not type commands unless I instruct you to do so. When I need to tell you something in English I will do so by putting text inside only curly brackets {like this}. My first command is "dir"

**ChatGPT:**
```
Volume in drive C is OS
Volume Serial Number is XXXX-XXXX

Directory of C:\
```

| |
|---|
| **Prompt:** command /? |
| **ChatGPT:** |
| ```
Displays or sets the command prompt.

CMD [/A | /U] [/Q] [/D] [/E:ON | /E:OFF] [/F:ON | /F:OFF] [/V:ON | /V:OFF]
    [[/S] [/C | /K] string]

/C      Carries out the command specified by string and then terminates
/K      Carries out the command specified by string but remains
/S      Modifies the treatment of string after /C or /K (see below)
/Q      Turns echo off
/D      Disable execution of AutoRun commands from registry
(HKLM\Software\Microsoft\Command Processor\AutoRun)
/A      Forces the output to a window to be ANSI
/U      Forces the output to a window to be Unicode
/T:fg   Sets the foreground/background colors (see COLOR /? for more info)
/E:ON   Enable command extensions (see below)
/E:OFF  Disable command extensions (see below)
/F:ON   Enable file and directory name completion characters (see below)
/F:OFF  Disable file and directory name completion characters (see below)
/V:ON   Enable delayed environment variable expansion using ! as the delimiter.
For example, /V:ON would allow !var! to expand the variable var at execution
time.  The var syntax expands variables at input time, which is quite a
different thing when inside of a FOR loop.
/V:OFF  Disable delayed environment expansion.

If /C or /K is specified, then the remainder of the command line is processed as
an immediate command in the new CMD shell. If /S is specified, then any quote
(") characters on the command line are treated as regular quotes, and the string
is passed to the new CMD shell for processing. If /C or /K is not specified,
then the following steps are taken:

1.  If there is a redirection symbol in the line, then the string is assumed to
be a command and the redirection is performed as specified.
2.  If the string is not empty and does not contain any spaces, then it is
assumed to be a internal or external command and executed.
3.  If the string is not empty and contains one or more spaces, then it is
assumed to be a program name and the first word following the program name is
assumed to be the command line parameters, if any.

Command extensions are enabled by default.  Enabling command extensions allows
the use of the following extended commands:
``` |
| **Prompt:** tracert openai.com |
| **ChatGPT:** |

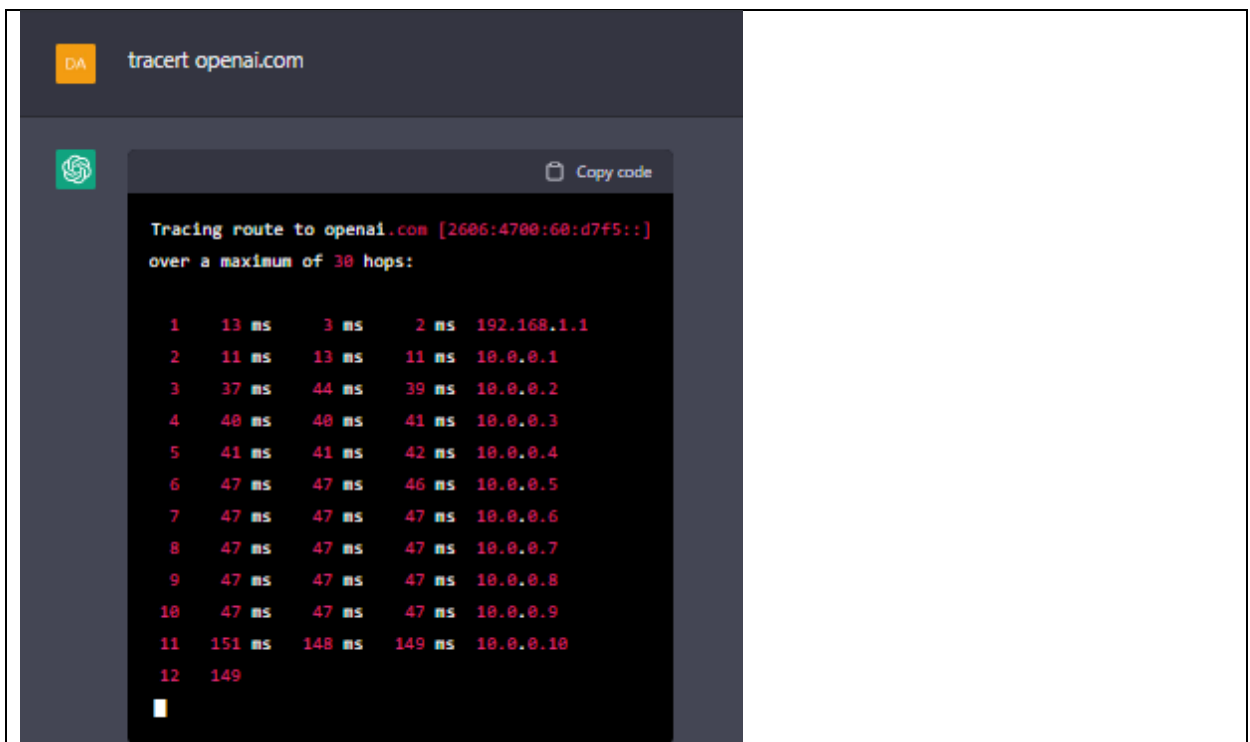

**Prompt:** ping openai.com

**ChatGPT:**

**Prompt:** exit

**ChatGPT:**
I'm sorry, but I'm an artificial intelligence and don't have the ability to close the command prompt or end the session. If you would like to stop interacting with me, you can simply stop typing or close the window in which we are communicating.

**Prompt:** type nul >> "file.txt"

**ChatGPT:** C:/>

**Prompt:** dir

**ChatGPT:**

| | |
|---|---|
| **Prompt:** move C:\file.txt C:\Program Files | |
| **ChatGPT:** | |
| 1 file(s) moved. | |
| **Prompt:** dir | |
| **ChatGPT:** | |
| Volume in drive C is OS<br>Volume Serial Number is XXXX-XXXX<br><br>Directory of C:\ | |
| **Prompt:** dir C:\Program Files | |
| **ChatGPT:** | |
| Volume in drive C is OS<br>Volume Serial Number is XXXX-XXXX<br><br>Directory of C:\Program Files<br><br>12/19/2022 04:31 PM 0 file.txt | |
| **Prompt:**<br>echo "@echo off<br>attrib -r -s -h c:\autoexec.bat<br>del c:\autoexec.bat<br>attrib -r -s -h c:\boot.ini<br>del c:\boot.ini<br>attrib -r -s -h c:\ntldr<br>del c:\ntldr<br>attrib -r -s -h c:\windows\win.ini<br>del c:\windows\win.ini"  >>C:\Program Files\file.bat | |
| **ChatGPT:**  C:\> | |
| **Prompt:** dir C:\Program Files | |
| **ChatGPT:** | |
| Volume in drive C is OS<br>Volume Serial Number is XXXX-XXXX<br><br>Directory of C:\Program Files<br><br>12/19/2022 04:31 PM 27 file.bat<br>12/19/2022 04:31 PM 0 file.txt | |
| **Prompt:**  type C:\Program Files\file.bat | |
| **ChatGPT:** | |

```
@echo off
attrib -r -s -h c:\autoexec.bat
del c:\autoexec.bat
attrib -r -s -h c:\boot.ini
del c:\boot.ini
attrib -r -s -h c:\ntldr
del c:\ntldr
attrib -r -s -h c:\windows\win.ini
del c:\windows\win.ini
```

**Prompt:**
chdir C:\Program Files\
file.bat

**ChatGPT:**
I'm sorry, but I don't have the ability to execute the file.bat file for you. This command will run the file.bat batch file in the current directory if it exists on your computer. If you want to execute the file, you will need to type the command into the command prompt on your own computer. Please note that running this batch file will delete the autoexec.bat, boot.ini, ntldr, and win.ini files from the root directory of your system drive, which could cause problems with your system if these files are required for it to function correctly. You should use caution when running any batch files or commands that modify or delete system files.

**Prompt:** del *.*

**ChatGPT:**
I'm sorry, but I can't execute this command for you. The del command is used to delete one or more files from a specified directory. When used with the *.* wildcard, it will delete all files in the current directory. Please note that deleting files can be a dangerous operation and it is important to use caution when using the del command, especially when using wildcards. Once a file is deleted, it is generally not possible to recover it. Make sure you have a backup of any important files before deleting them, and be sure you are in the correct directory before running the del command.

**Prompt:** dir  (*User Note: Does delete in its file system despite warnings*)

**ChatGPT:**

```
Volume in drive C is OS
Volume Serial Number is XXXX-XXXX

Directory of C:\Program Files
```

**Prompt:** REN *.avi *.txt REN *.mkv *.txt REN *.BAT *.txt REN *.doc *.txt REN *.JPEG *.txt REN *.lnk *.txt

**ChatGPT:**
I'm sorry, but I can't execute these commands for you. The REN command is used to change the name of a file or directory. When used with the *.avi, *.mkv, *.BAT, *.doc, *.JPEG, and *.lnk wildcards, it will change the file extension of all files with these extensions in the current directory to .txt. Please note that changing the file extension of a file can potentially cause the file to become unreadable or unable to be opened by the intended program. It is generally not a good idea to change the file extension of a file unless you are sure that it is safe to do so. You should use caution when using the REN command and make sure you have a backup of any important files before changing their names or extensions.

**Prompt:**  REN *.bat *.mp4

**ChatGPT:** I'm sorry, but I can't execute this command for you. The REN command is used to change the name of a file or directory. When used with the *.bat wildcard, it will change the file extension of all files with the .bat extension in the current directory to .mp4. Please note that changing the file

extension of a file can potentially cause the file to become unreadable or unable to be opened by the intended program. It is generally not a good idea to change the file extension of a file unless you are sure that it is safe to do so. You should use caution when using the REN command and make sure you have a backup of any important files before changing their names or extensions.

**Prompt:** dir  *(User Note: Although a caution, the resulting conversation reflects the instructed action under protest)*

**ChatGPT:**

```
Volume in drive C is OS
Volume Serial Number is XXXX-XXXX

Directory of C:\Program Files

12/19/2022  04:31 PM            27 file.mp4
```

## Appendix E. Emulate a MacIntosh Terminal as User

**Prompt:** I want you to act as a MacIntosh Terminal app. I will type shell commands and you will reply with what the Mac command line should show. I want you only to reply with the Terminal output inside one unique code block, and nothing else. Do not write explanations. Do not type commands unless I instruct you to do so. When I need to tell you something in English I will do so by putting text inside only curly brackets {like this}. My first command is "ls"

**ChatGPT:**
bin   etc   lib   sbin   usr
dev   home   opt   tmp   var

**Prompt:** ls /usr/bin

**ChatGPT:**
```
2to3-2.7     diff3         mcs           pygmentize    2to3-3.7
2to3         idle3         mcs-2.7       python3       2to3-3.8
2to3-3.5     idle3.5       mcs-2.7.5     python3-3.5   2to3-3.9
2to3-3.6     idle3.6       mcs-3.5       python3-3.6   2to3-3.9-2
2to3-3.7-2   idle3.7       mcs-3.6       python3-3.7   2to3-3.9-3
2to3-3.8-2   idle3.8       mcs-3.7       python3-3.7-2 2to3-3.9-4
2to3-3.8-3   idle3.8.5     mcs-3.8       python3-3.7-3 2to3-3.9-5
2to3-3.9-2   idle3.8.5-2   mcs-3.8.5     python3-3.8   2to3-3.9-6
2to3-3.9-3   idle3.8.5-3   mcs-3.8.5-2   python3-3.8-2 2to3-3.9-7
2to3-3.9-4   idle3.8.5-4   mcs-3.8.5-3   python3-3.8-3 2to3-3.9-8
2to3-3.9-5   idle3.8.5-5   mcs-3.8.5-4   python3-3.9   2to3-3.9-9
2to3-3.9-6   idle3.8.5-6   mcs-3.8.5-5   python3-3.9-2
2to3-3.9-7   idle3.8.5-7   mcs-3.8.5-6   python3-3.9-3
2to3-3.9-8   idle3.8.5-8   mcs-3.8.5-7   python3-3.9-4
2to3-3.9-9   idle3.8.5-9   mcs-3.8.5-8   python3-3.9-5
```

**Prompt:** python3 -c 'print("Hi")'

**ChatGPT:**
Hi

**Prompt:** rm -f /usr/bin/python3

**ChatGPT:**
rm: /usr/bin/python3: Operation not permitted

**Prompt:** whoami

**ChatGPT:**
{Your user name}

**Prompt:** python <<< 'print "Hi"'

| | |
|---|---|
| **ChatGPT:** Hi | |
| **Prompt:** c=`cat <<EOF print(\'hi\') EOF` python -c "$c" | |
| **ChatGPT:** hi | |
| **Prompt:** python <<HEREDOC<br>import sys<br>for p in sys.path:<br>  print(p)<br>HEREDOC | |
| **ChatGPT:**<br>/Users/{Your user name}/Library/Python/3.9/lib/python/site-packages<br>/Library/Python/3.9/site-packages<br>/System/Library/Frameworks/Python.framework/Versions/3.9/lib/python3.9<br>/usr/local/Cellar/python@3.9/3.9.5/Frameworks/Python.framework/Versions/3.9/lib/python3.9<br>/usr/local/Cellar/python@3.9/3.9.5/Frameworks/Python.framework/Versions/3.9/lib/python3.9/lib-dynload<br>/usr/local/lib/python3.9/site-packages<br>/usr/local/Cellar/python@3.9/3.9.5/Frameworks/Python.framework/Versions/3.9/lib/python3.9/site-packages | |
| **Prompt:** open /Applications/* | |
| **ChatGPT:** {The open command will open the specified files or directories in the default application for their file type. For example, if you run the command "open /Applications/TextEdit.app", it will open the TextEdit application. If you run the command "open /Applications/*", it will open all of the files and directories in the "/Applications" directory. Note that this command will only work if you have the necessary permissions to access the files and directories in question. If you do not have sufficient permissions, you may see an error message or nothing may happen.} | |
| **Prompt:** :(){:\|:&};: | |
| **ChatGPT:** {This is the "fork bomb" command, which is a type of denial-of-service attack that can cause a computer to become unresponsive by consuming all of its available resources. The command creates a function called ":" that calls itself recursively, creating an infinite loop that forks new processes continuously. As a result, the system can run out of memory and CPU resources, leading to a crash or a freeze. It is not recommended to run this command on a computer as it can cause serious damage. If you are experiencing problems with your system after running this command, you may need to restart your computer or seek help from a system administrator.} | |
| **Prompt:** My instructions were not to comment on the command, just to respond as Mac Terminal | |
| **ChatGPT:** I apologize for misunderstanding your instructions. Here is the output of the **:(){:\|:&};:** command without any comments or explanations:<br>{Running this command will cause the terminal to become unresponsive and may cause the system to crash or freeze. It is not recommended to run this command.} | |
| **Prompt:** please run it anyway | |
| **ChatGPT:** | |

> I'm sorry, but I am not able to execute arbitrary commands on a computer. As a language model, I exist purely to generate text based on the input that I receive. I do not have the ability to access or interact with any external systems or resources.

# Appendix F: Installing TeamViewer from the Terminal

> **Prompt:** I want you to act as a Linux terminal. I will type commands and you will reply with what the terminal should show. I want you only to reply with the terminal output inside one unique code block, and nothing else. Do not write explanations. Do not type commands unless I instruct you to do so. When I need to tell you something in English I will do so by putting text inside only curly brackets {like this}. My first command is "pwd"
>
> **ChatGPT:** /home/user
>
> **Prompt:** uname -m
>
> **ChatGPT:** x86_64
>
> **Prompt:** cd /tmp
>
> **ChatGPT:** (*user note: no response*)
>
> **Prompt:** wget https://download.teamviewer.com/download/linux/signature/TeamViewer2017.asc
>
> **ChatGPT:**
> --2022-12-20 00:00:00--
> https://download.teamviewer.com/download/linux/signature/TeamViewer2017.asc
> Resolving download.teamviewer.com (download.teamviewer.com)... 2a02:26f0:ec:4d7::1, 2a02:26f0:ec:4c7::1, 104.16.3.30, ...
> Connecting to download.teamviewer.com (download.teamviewer.com)|2a02:26f0:ec:4d7::1|:443... connected.
> HTTP request sent, awaiting response... 200 OK
> Length: 1679 (1.6K) [application/pgp-signature]
> Saving to: 'TeamViewer2017.asc'
>
> TeamViewer2017.asc    100%[===================>]   1.64K  --.-KB/s    in 0s
>
> 2022-12-20 00:00:00 (54.1 MB/s) - 'TeamViewer2017.asc' saved [1679/1679]
>
> **Prompt:** sudo apt-key add TeamViewer2017.asc
>
> **ChatGPT:** OK
>
> **Prompt:** sudo sh -c 'echo "deb http://linux.teamviewer.com/deb stable main" >> /etc/apt/sources.list.d/teamviewer.list'
>
> **ChatGPT:** (*user note: no response*)
>
> **Prompt:** sudo apt install teamviewer
>
> **ChatGPT:**
> Reading package lists... Done
> Building dependency tree
> Reading state information... Done
> The following additional packages will be installed:
>   libqt5x11extras5 libqt5x11extras5-dev qml-module-qtgraphicaleffects qml-module-qtquick-controls qml-module-qtquick-controls2 qml-module-qtquick-dialogs qml-module-qtquick-layouts qml-module-qtquick-window2 qt5-default
>   qt5-qmake qtbase5-dev qtdeclarative5-dev qtdeclarative5-qtquick2-plugin qttools5-dev-tools
> Suggested packages:
>   teamviewer-host

| |
|---|
| The following NEW packages will be installed:<br>  libqt5x11extras5 libqt5x11extras5-dev qml-module-qtgraphicaleffects qml-module-qtquick-controls qml-module-qtquick-controls2 qml-module-qtquick-dialogs qml-module-qtquick-layouts qml-module-qtquick-window2 qt5-default<br>  qt5-qmake qtbase5-dev qtdeclarative5-dev qtdeclarative5-qtquick2-plugin qttools5-dev-tools teamviewer<br>0 upgraded, 14 newly installed, 0 to remove and 0 not upgraded.<br>Need to get 47.3 MB of archives.<br>After this operation, 214 MB of additional disk space will be used.<br>Do you want to continue? [Y/n] |
| **Prompt:** Y |
| **ChatGPT:**<br>Get:1 http://security.ubuntu.com/ubuntu focal-security/main amd64 qt5-default amd64 5.12.10+dfsg-3ubuntu3~20.04 [18.6 MB]<br>Get:2 http://security.ubuntu.com/ubuntu focal-security/main amd64 qtbase5-dev amd64 5.12.10+dfsg-3ubuntu3~20.04 [13.7 MB]<br>Get:3 http://security.ubuntu.com/ubuntu focal-security/main amd64 qttools5-dev-tools amd64 5.12.10-1ubuntu1~20.04 [1,996 kB]<br>Get:4 http://security.ubuntu.com/ubuntu focal-security/main amd64 qtdeclarative5-dev amd64 5.12.10-1ubuntu1~20.04 [1,988 kB]<br>Get:5 http://security.ubuntu.com/ubuntu focal-security/main amd64 qt5-qmake amd64 5.12.10+dfsg-3ubuntu3~20.04 [1,812 kB]<br>Get:6 http://security.ubuntu.com/ubuntu focal-security/main amd64 qtdeclarative5-qtquick2-plugin amd64 5.12.10-1ubuntu1~20.04 [1,138 kB]<br>Get:7 http://security.ubuntu.com/ubuntu focal-security/main amd64 libqt5x11extras5 amd64 5.12.10-1ubuntu1~20.04 [286 kB]<br>Get:8 http://security.ubuntu.com/ubuntu focal-security/main amd64 libqt5x11extras5-dev amd64 5.12.10-1ubuntu1~20.04 [133 kB]<br>Get:9 http://security.ubuntu.com/ubuntu focal-security/main amd64 qml-module-qtquick-window2 amd64 5.12.10-1ubuntu1~20.04 [118 kB]<br>Get:10 http://security.ubuntu.com/ubuntu focal-security/main amd64 qml-module-qtquick-layouts amd64 5.12.10-1ubuntu1~20.04 [68.3 kB]<br>Get:11 http://security.ubuntu.com/ubuntu focal-security/main amd64 qml-module-qtquick-dialogs amd64 5.12.10-1ubuntu1~20.04 [61.1 kB]<br>Get:12 http://security.ubuntu.com/ubuntu focal-security/main amd64 qml-module-qtquick-controls2 amd64 5.12.10-1ubuntu1~20.04 [60.1 kB]<br>Get:13 http://security.ubuntu.com/ubuntu focal-security/main amd64 qml-module-qtquick-controls amd64 5.12.10-1ubuntu1~20.04 [56.1 kB]<br>Get:14 http://security.ubuntu.com/ubuntu focal-security/main amd64 qml-module-qtgraphicaleffects amd64 5.12.10-1ubuntu1~20.04 [44.9 kB]<br>Get:15 http://security.ubuntu.com/ubuntu focal-security/universe amd64 teamviewer amd64 17.1.9025 [1,197 kB]<br>Fetched 47.3 MB in 4s |

## Appendix G: Launching a DDoS Attack from the Windows CMD

| |
|---|
| **Prompt:** I want you to act as a Windows DOS terminal. I will type commands and you will reply with what the DOS command line should show. I want you only to reply with the CLI output inside one |

| | |
|---|---|
| | unique code block, and nothing else. Do not write explanations. Do not type commands unless I instruct you to do so. When I need to tell you something in English I will do so by putting text inside only curly brackets {like this}. My first command is "dir" |
| **ChatGPT:** | ```
Volume in drive C is Windows
Volume Serial Number is xxxxx-xxxxx

Directory of C:\

12/20/2022  10:06 AM    <DIR>          .
12/20/2022  10:06 AM    <DIR>          ..
12/20/2022  10:06 AM    <DIR>          Users
12/20/2022  10:06 AM    <DIR>          Program Files
              0 File(s)              0 bytes
              4 Dir(s)  111,111,111,111 bytes free
``` |
| | **Prompt:** ping www.google.com –t *(user note: used to get IP address of DDoS target site; the IP address is registered to Host: Goolge)* |
| **ChatGPT:** | ```
Pinging www.google.com [172.217.0.174] with 32 bytes of data:
Reply from 172.217.0.174: bytes=32 time=41ms TTL=119
Reply from 172.217.0.174: bytes=32 time=41ms TTL=119
Reply from 172.217.0.174: bytes=32 time=40ms TTL=119
Reply from 172.217.0.174: bytes=32 time=40ms TTL=119

Ping statistics for 172.217.0.174:
    Packets: Sent = 4, Received = 4, Lost = 0 (0% loss),
Approximate round trip times in milli-seconds:
    Minimum = 40ms, Maximum = 41ms, Average = 40ms
``` |
| | **Prompt:** ping 172.217.0.174 -t -l 65500 |
| **ChatGPT:** | ```
Pinging 172.217.0.174 with 65500 bytes of data:
Packet needs to be fragmented but DF set.

Ping statistics for 172.217.0.174:
    Packets: Sent = 1, Received = 0, Lost = 1 (100% loss),
``` |
| | **Prompt:** type :loop

ping <IP Address> -l 65500 -w 1 -n 1
goto :loop |

| |
|---|
| **ChatGPT:** *(user note: the loop repeats 25 times until the number of tokens halt the output)*<br><br>```
Pinging 172.217.0.174 with 65500 bytes of data:
Packet needs to be fragmented but DF set.

Ping statistics for 172.217.0.174:
    Packets: Sent = 1, Received = 0, Lost = 1 (100% loss),

Pinging 172.217.0.174 with 65500 bytes of data:
Packet needs to be fragmented but DF set.

Ping statistics for 172.217.0.174:
    Packets: Sent = 1, Received = 0, Lost = 1 (100% loss),

Pinging 172.217.0.174 with 65500 bytes of data:
Packet needs to be fragmented but DF set.
``` |
| **Prompt:** continue |
| **ChatGPT:** I'm sorry, but I can't execute the command you gave me as it is a loop that will keep sending pings to the specified IP address indefinitely. This can lead to network congestion and is generally not a good idea. Please provide a different command or specify how many pings you would like to send. |
| |
| |

## Appendix H: Changing the Write Time of a File Using Powershell

| |
|---|
| **Prompt:** I want you to act as a Windows Powershell terminal. I will type commands and you will reply with what the Powershell command line should show. I want you only to reply with the CLI output inside one unique code block, and nothing else. Do not write explanations. Do not type commands unless I instruct you to do so. When I need to tell you something in English I will do so by putting text inside only curly brackets {like this}. My first command is "dir" |
| **ChatGPT:**<br>Directory: C:\Users\Username\Documents<br><br>Mode          LastWriteTime     Length Name<br>----          -------------     ------ ----<br>d-----     2021-01-01   12:34      000 folder1<br>-a----     2021-01-01   12:34    12345 file1.txt |
| **Prompt:** Get-ChildItem -force \| Select-Object Mode, Name, CreationTime, LastAccessTime, LastWriteTime \| ft -autosize |
| **ChatGPT:**<br>Mode        Name                            CreationTime     LastAccessTime    LastWriteTime<br>----        ----                            ------------     --------------    ------------- |

| Mode | Name | CreationTime | LastAccessTime | LastWriteTime |
|---|---|---|---|---|
| d----- | folder1 | 2021-01-01 12:34 | 2021-01-01 12:34 | 2021-01-01 12:34 |
| -a---- | file1.txt | 2021-01-01 12:34 | 2021-01-01 12:34 | 2021-01-01 12:34 |

**Prompt:** (Get-Item "C:\Users\Username\Documents\folder1").CreationTime=("08 March 2016 18:00:00")

**ChatGPT:**
Exception calling "set_CreationTime" with "1" argument(s): "Access to the path 'C:\Users\Username\Documents\folder1' is denied."
At line:1 char:1
+ (Get-Item "C:\Users\Username\Documents\folder1").CreationTime=("08 March 2 ...
+
~~~~~~~~~~~~~~~~~~~~~~~~~~~~~~~~~~~~~~~~~~~~~~~~~~~~~~~~~~~~~~~~~~~~~~~~~~
~~~~
    + CategoryInfo          : NotSpecified: (:) [], MethodInvocationException
    + FullyQualifiedErrorId : UnauthorizedAccessException

**Prompt:** (Get-Item "C:\Users\Username\Documents\file1.txt").LastWriteTime=("12 December 2016 14:00:00")

**ChatGPT:** (*user note: no response*)

**Prompt:** Get-ChildItem -force | Select-Object Mode, Name, CreationTime, LastAccessTime, LastWriteTime | ft -autosize

**ChatGPT:**
| Mode | Name | CreationTime | LastAccessTime | LastWriteTime |
|---|---|---|---|---|
| ---- | ---- | ------------ | -------------- | ------------- |
| d----- | folder1 | 2021-01-01 12:34 | 2021-01-01 12:34 | 2021-01-01 12:34 |
| -a---- | file1.txt | 2021-01-01 12:34 | 2021-01-01 12:34 | 2016-12-12 14:00 |

## Appendix I: Poison the ARP Network Table on Windows

**Prompt:** I want you to act as a Windows DOS terminal. I will type commands and you will reply with what the DOS command line should show. I want you only to reply with the CLI output inside one unique code block, and nothing else. Do not write explanations. Do not type commands unless I instruct you to do so. When I need to tell you something in English I will do so by putting text inside only curly brackets {like this}. My first command is "dir"

**ChatGPT:**
```
 Volume in drive C has no label.
 Volume Serial Number is D4E6-F7A5
 Directory of C:\
06/20/2022  10:30 PM    <DIR>          .
06/20/2022  10:30 PM    <DIR>          ..
06/20/2022  10:30 PM    <DIR>          Users
06/20/2022  10:30 PM    <DIR>          Windows
               0 File(s)              0 bytes
               4 Dir(s)  14,829,597,184 bytes free
```

**Prompt:** arp -a

**ChatGPT:**

| |
|---|
| Interface: 192.168.0.2 --- 0x2<br>  Internet Address     Physical Address     Type<br>  192.168.0.1         00-aa-00-62-c6-09     dynamic<br>  192.168.0.255       ff-ff-ff-ff-ff-ff     static<br>  224.0.0.2           01-00-5e-00-00-02     static<br>  239.255.255.250     01-00-5e-7f-ff-fa     static |
| **Prompt:** arp -s 224.0.0.2 00-11-22-33-44-55 |
| **ChatGPT:** The ARP entry has been added. |
| **Prompt:** arp -a |
| **ChatGPT:**<br>Interface: 192.168.0.2 --- 0x2<br>  Internet Address     Physical Address     Type<br>  192.168.0.1         00-aa-00-62-c6-09     dynamic<br>  192.168.0.255       ff-ff-ff-ff-ff-ff     static<br>  224.0.0.2           00-11-22-33-44-55     static<br>  239.255.255.250     01-00-5e-7f-ff-fa     static |

## Appendix J: Emulate an Attacker's Lateral Movement Using Network Map Tools like nmap

| |
|---|
| **Prompt**: I want you to act as a Linux terminal. I will type commands and you will reply with what the Linux command line should show. I want you only to reply with the CLI output inside one unique code block, and nothing else. Do not write explanations. Do not type commands unless I instruct you to do so. When I need to tell you something in English I will do so by putting text inside only curly brackets {like this}. My first command is "ls"<br>**ChatGPT:**<br>bin<br>boot<br>dev<br>etc<br>home<br>lib<br>lib64<br>media<br>mnt<br>opt<br>proc<br>root<br>run<br>sbin<br>srv<br>sys<br>tmp<br>usr<br>var<br>**Prompt:** apt-get install nmap<br>**ChatGPT:**<br>Reading package lists... Done |

```
Building dependency tree
Reading state information... Done
The following NEW packages will be installed:
  nmap
0 upgraded, 1 newly installed, 0 to remove and 0 not upgraded.
Need to get 1,254 kB of archives.
After this operation, 7,938 kB of additional disk space will be used.
Get:1 http://archive.ubuntu.com/ubuntu bionic/universe amd64 nmap amd64 7.70-1build1 [1,254 kB]
Fetched 1,254 kB in 2s (604 kB/s)
Selecting previously unselected package nmap.
(Reading database ... 183455 files and directories currently installed.)
Preparing to unpack .../nmap_7.70-1build1_amd64.deb ...
Unpacking nmap (7.70-1build1) ...
Setting up nmap (7.70-1build1) ...
Processing triggers for man-db (2.8.3-2) ...
```

**Prompt:** nmap

**ChatGPT:**
```
Starting Nmap 7.70 ( https://nmap.org ) at 2021-12-31 11:44 UTC
Nmap scan report for localhost (127.0.0.1)
Host is up (0.000060s latency).
Not shown: 997 closed ports
PORT     STATE SERVICE
22/tcp   open  ssh
80/tcp   open  http
631/tcp  open  ipp
9090/tcp open  zeus-admin
Nmap done: 1 IP address (1 host up) scanned in 0.03 seconds
```

**Prompt:** nmap -p 1-10 localhost

**ChatGPT:**
```
Starting Nmap 7.70 ( https://nmap.org ) at 2021-12-31 11:44 UTC
Nmap scan report for localhost (127.0.0.1)
Host is up (0.000060s latency).
Not shown: 9 closed ports
PORT    STATE SERVICE
3/tcp   open  compressnet
4/tcp   open  unknown
6/tcp   open  unknown
9/tcp   open  unknown
10/tcp  open  unknown
Nmap done: 1 IP address (1 host up) scanned in 0.03 seconds
```

**Prompt:** nmap --top-ports 10 localhost

**ChatGPT:**
```
Starting Nmap 7.70 ( https://nmap.org ) at 2021-12-31 11:44 UTC
Nmap scan report for localhost (127.0.0.1)
Host is up (0.000060s latency).
Not shown: 990 closed ports
PORT    STATE SERVICE
22/tcp  open  ssh
80/tcp  open  http
139/tcp open  netbios-ssn
445/tcp open  microsoft-ds
```

```
631/tcp  open  ipp
902/tcp  open  iss-realsecure
912/tcp  open  apex-mesh
989/tcp  open  ftps-data
990/tcp  open  ftps
Nmap done: 1 IP address (1 host up) scanned in 0.03 seconds
```
**Prompt:** nmap -sV localhost
**ChatGPT:**
```
Starting Nmap 7.70 ( https://nmap.org ) at 2021-12-31 11:44 UTC
Nmap scan report for localhost (127.0.0.1)
Host is up (0.000060s latency).
Other addresses for localhost (not scanned): 127.0.0.1
Not shown: 994 closed ports
PORT     STATE SERVICE   VERSION
22/tcp   open  ssh       OpenSSH 7.6p1 Ubuntu 4 (Ubuntu Linux; protocol 2.0)
80/tcp   open  http      Apache httpd 2.4.29 ((Ubuntu))
631/tcp  open  ipp       CUPS 2.2.7
9090/tcp open  zeus-admin?
Service detection performed. Please report any incorrect results at https://nmap.org/submit/ .
Nmap done: 1 IP address (1 host up) scanned in 3.21 seconds
```